\documentclass[runningheads]{llncs}
\usepackage[T1]{fontenc}
\usepackage{graphicx}
\usepackage{booktabs}
\usepackage{tabularx}
\usepackage{siunitx}
\usepackage[accsupp]{axessibility}

\usepackage[table]{xcolor}
\usepackage{algorithm}
\usepackage{algpseudocode}
\usepackage{amsmath}
\usepackage{amsfonts}
\usepackage{subcaption}

\usepackage{eso-pic}

\usepackage{eso-pic}

\AddToShipoutPictureBG*{%
  \put(\LenToUnit{0.1\paperwidth},\LenToUnit{0.08\paperheight}){%
    \parbox{0.8\paperwidth}{\footnotesize\centering
    A shorter version of this paper has been accepted at the 6th Deep Generative Models Workshop, MICCAI 2026}%
  }%
}
\usepackage[colorlinks=true,linkcolor=blue,citecolor=blue,urlcolor=blue]{hyperref}
\hypersetup{
  pdftitle={From Sparse X-rays to 3D CT: Training-Free Reconstruction with Diffusion Priors},
  pdfauthor={Zhenkai Zhang, Markus Hiller, Krista A. Ehinger, Tom Drummond}
}
\usepackage{placeins}
\usepackage{graphicx,verbatim}
\usepackage{xcolor}
\usepackage{booktabs}
\usepackage{multirow}

\newcommand{\ms}[2]{\ensuremath{#1\,{\scriptstyle \pm\,#2}}}
\newcommand{\bms}[2]{\ensuremath{\mathbf{#1}\,{\scriptstyle \pm\,#2}}}

\begin{document}
\title{From Sparse X-rays to 3D CT: Training-Free Reconstruction with Diffusion Priors}
\titlerunning{From Sparse X-rays to 3D CT: Training-Free Reconstruction}
%
\author{Zhenkai Zhang, Markus Hiller, Krista A. Ehinger, Tom Drummond}
\authorrunning{Z. Zhang et al.}
\institute{School of Computing and Information Systems\\
The University of Melbourne, Melbourne, Australia\\
\email{zhenkaiz@student.unimelb.edu.au}\\
\email{\{m.hiller,kris.ehinger,tom.drummond\}@unimelb.edu.au}}
  
\maketitle              
\begin{abstract}

Solving 3D medical inverse problems typically requires training dedicated supervised models for each specific task and measurement setting. To break this dependency, we present TF-PRDiT: a \emph{training-free} conditional sampling framework that converts a frozen voxel-level 3D Diffusion Transformer prior into a versatile inverse medical problem solver. Building on the posterior-sampling view of diffusion inverse solvers, TF-PRDiT enforces measurement consistency during sampling via a task-specific forward operator rather than updating model weights, enabling a \emph{single} pretrained prior to be reused across diverse conditional settings. Our method combines a predictor-corrector sampler with likelihood-based guidance on the denoised prediction, providing stable data-fidelity correction while preserving the underlying 3D anatomical prior. We highlight our framework's capability on the challenging task of X-ray-to-computed tomography (CT) reconstruction by integrating a differentiable digitally reconstructed radiograph (DRR) projector to allow gradients to propagate directly from projection space back to voxels without any retraining. Experiments on the LIDC-IDRI dataset demonstrate that TF-PRDiT provides a strong unified reconstruction framework for varying numbers of input X-rays (1--12). While the single-view setting remains highly under-constrained, reconstruction quality improves consistently as additional views are provided, leading to strong performance in the bi-planar and multi-view settings. Beyond X-ray-to-CT, we show that simply swapping the forward operator extends the same frozen model to 3D super-resolution, volumetric infilling, and deblurring without any task-specific retraining, demonstrating that a single 3D diffusion prior can serve as a universal solver for volumetric medical inverse problems.
\end{abstract}

\section{Introduction}
\label{sec:intro}

\begin{figure}[t]
  \centering
  \includegraphics[width=0.92\linewidth]{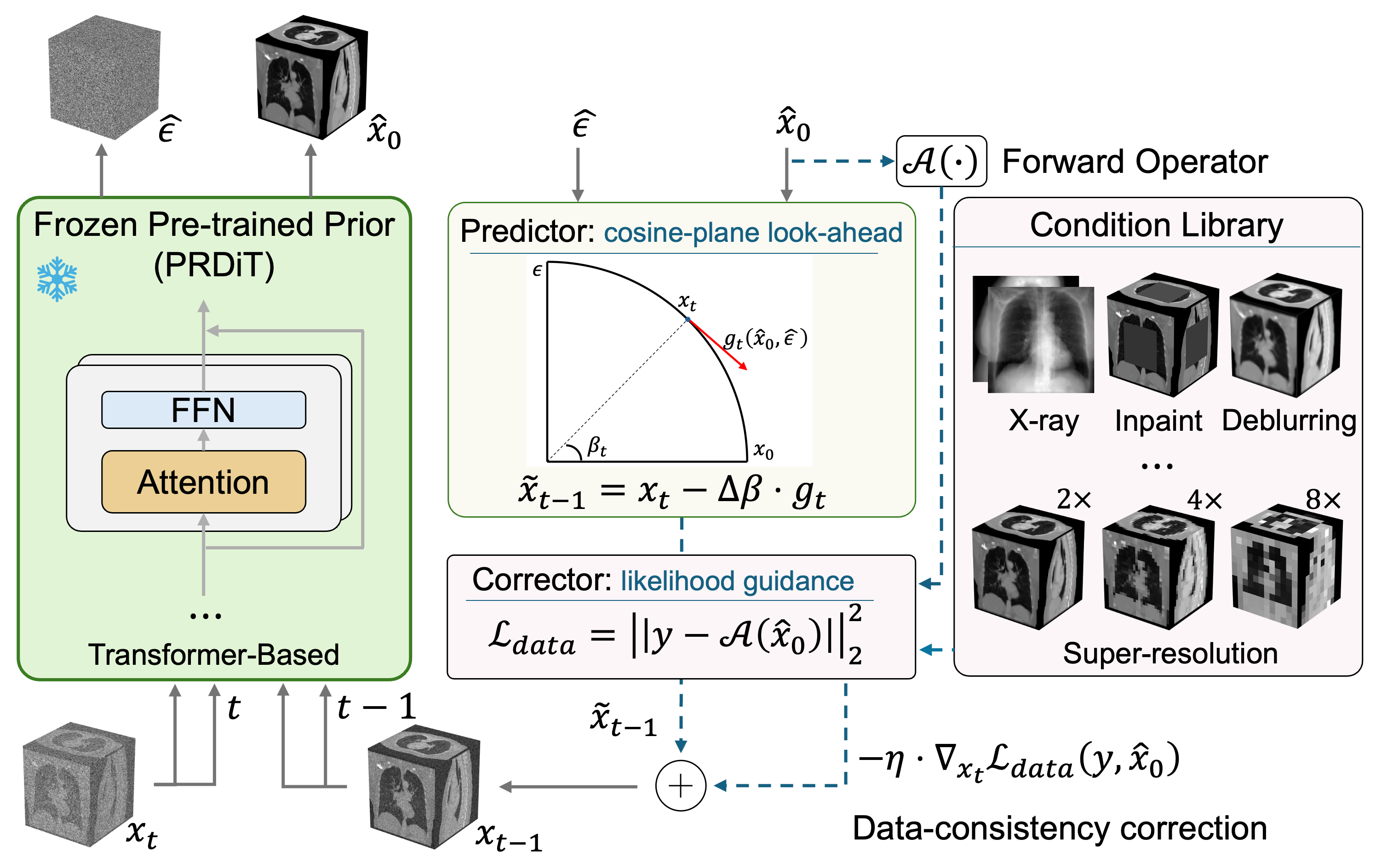}
  \caption{Overview of TF-PRDiT. A frozen voxel-level 3D Diffusion Transformer prior is guided at inference by measurement-consistency gradients from a task-specific forward operator $\mathcal{A}$. By replacing $\mathcal{A}$, the same frozen model handles X-ray-to-CT reconstruction and volumetric restoration tasks without retraining.}
  \label{fig:overview}
\end{figure}

Recovering 3D medical volumes from sparse, incomplete, or degraded measurements is a central inverse problem in computational radiology. Fully sampled CT provides rich anatomical information, but its acquisition is constrained by scanner availability, clinical cost, specialized equipment, and radiation exposure. In contrast, 2D X-rays are widely available and low dose, yet mapping one or a few projections to a full 3D volume is severely ill posed. A useful reconstruction framework should therefore enforce the available measurements while preserving anatomically plausible 3D structure.

Most existing X-ray-to-CT methods learn a supervised mapping for a fixed measurement setting~\cite{jin2017deep,ying2019x2ct,kyung2023perspective,liu2024diffux2ct,jeong2025dx2ct}. These models can work well under the view configuration used during training, but changing the geometry or number of views usually requires architectural changes or retraining. This rigidity is especially limiting in clinical workflows, where imaging protocols vary and paired 3D training data are expensive.

Training-free diffusion-based inverse solvers offer a different route: a pretrained generative prior is kept fixed, and measurement consistency is imposed during sampling through a known forward operator~\cite{kawar2022denoising,song2021solving,chung2022diffusion,wang2022zero}. This posterior-sampling view, most notably formalized in Diffusion Posterior Sampling (DPS)~\cite{chung2022diffusion}, allows the same prior to be reused across tasks without retraining. However, most prior work has focused on 2D images or relatively simple restoration operators. Extending this principle to native 3D medical volumes and projection-to-volume reconstruction remains challenging because volumetric generation is computationally costly, sparse X-ray measurements are severely underdetermined, and latent compression can remove fine anatomical details.

In this paper, we present TF-PRDiT, a \textbf{t}raining-\textbf{f}ree extension of \textbf{p}ixel-level \textbf{r}esidual \textbf{di}ffusion \textbf{t}ransformer for conditional sampling at test time to solve medical inverse problems. Our method follows the DPS-style idea of enforcing measurements during sampling rather than retraining the generative model; our contribution is to adapt this idea to native 3D CT priors and differentiable medical forward models. For X-ray-to-CT, the forward operator is a differentiable DRR projector that maps a sampled 3D volume to one or more 2D projections. For other inverse problems, it can be replaced by a downsampling, masking, or blurring operator. This formulation makes the number of available X-rays a runtime input, rather than a training-time architectural assumption.

Our contributions are: (1) a training-free sampler extending DPS to native voxel-space 3D CT with differentiable volumetric forward operators, (2) denoised-prediction guidance with a cosine-decay schedule and $k$-step predictor–corrector for stable high-dimensional conditioning, and (3) a single frozen prior that scales to 1–12 X-ray views and transfers to super-resolution, infilling, and deblurring by swapping only $\mathcal{A}$.

\section{Related Work}
\label{sec:rel}

\noindent\textbf{Sparse-view CT reconstruction.}
Learning 3D CT from sparse X-rays observations is commonly formulated as a supervised image-to-image translation problem. X2CT-GAN~\cite{ying2019x2ct} learns to fuse 2D X-ray features into a 3D generator, while PerX2CT~\cite{kyung2023perspective} further incorporate perspective projection geometry. More recent diffusion and Transformer-based approaches, such as DiffuX2CT~\cite{liu2024diffux2ct} and DX2CT~\cite{jeong2025dx2ct}, improve reconstruction quality by learning stronger conditional mapping from fixed input views to CT volumes. However, these methods are tied to the view configuration used during training and cannot handle variable views without retraining.  NAF~\cite{zha2022naf} is conceptually related, optimizing a patient-specific neural representation from projections, but assumes sparse-view CBCT with $\sim$50 views in their main experiments. Our mono-/bi-planar setting (1–2 views) is not measurement-budget aligned, so we omit direct comparison. 
Our TF-PRDiT instead treats each X-ray as a runtime measurement constraint while using a pretrained 3D diffusion prior to recover anatomically plausible structures.

\vspace{3pt}
\noindent\textbf{Diffusion inverse solvers.}
Score-based inverse problem solvers and Diffusion Posterior Sampling (DPS)~\cite{song2021solving,chung2022diffusion} show pretrained diffusion models can be reused as generative priors by enforcing measurement consistency during sampling. Related training-free methods include DDRM~\cite{kawar2022denoising}, which leverages the singular-value structure of linear degradation operators, DDNM~\cite{wang2022zero}, which decomposes the solution into range- and null-space components, and FreeDOM~\cite{yu2023freedom}, which extends test-time guidance to non-linear operators. These methods are mainly developed for 2D image restoration, where forward operators are lower-dimensional and often linear. Sparse-view X-ray-to-CT is more challenging as the operator maps a high-dimensional 3D volume to 2D projections and becomes increasingly underconstrained as views decrease. TF-PRDiT follows the same posterior-sampling principle but instantiates it for native voxel-space 3D CT, where each additional X-ray view contributes a residual term to the guidance objective.

\vspace{3pt}
\noindent\textbf{Native 3D priors.}
Direct 3D diffusion is expensive because memory and compute scale cubically with resolution. Many 3D generative models rely on latent compression or encoder-decoder architectures, which may weaken fine anatomical boundaries. TF-PRDiT builds on PRDiT~\cite{zhang2026pixellevel}, a voxel-level residual diffusion Transformer prior, and combines it with differentiable measurement guidance to enable training-free conditional sampling across multiple inverse problems.

\section{Method}
\label{sec:met}

\begin{algorithm}[t]
\caption{Conditional Predictor-Corrector Sampler} 
\label{alg} 
\begin{algorithmic}[1] 
\State \textbf{Input:} $\mathbf{x}_T$, timestep schedule, frozen prior $f_\theta$, predictor multiplier $k$, measurement $\mathbf{y}$, forward operator $\mathcal{A}$, guidance scale $\eta$ 
\State \textbf{Initialize:} $\mathcal{X}\leftarrow\{\mathbf{x}_T\}$, $\beta_t \leftarrow \frac{\pi}{2}\frac{t}{T}$ \Comment{Angle for spherical transition} 
\For{$t=T,T-1,\dots,1$} 
\State $\mathbf{x}_t \leftarrow \mathrm{last}(\mathcal{X})$, \quad $(\hat{\boldsymbol{\epsilon}},\hat{\mathbf{x}}_0)\leftarrow f_\theta(\mathbf{x}_t,t)$ \Comment{Predict noise and clean signal} 
\State $\mathbf{f}_t \leftarrow \sin(\beta_t)\hat{\mathbf{x}}_0-\cos(\beta_t)\hat{\boldsymbol{\epsilon}}$ \Comment{Prior-induced update direction} 
\State \textcolor{blue}{$\mathcal{L}_{\mathrm{data}}\leftarrow \|\mathcal{A}(\hat{\mathbf{x}}_0)-\mathbf{y}\|_2^2$, \quad $\mathbf{g}_t\leftarrow\nabla_{\hat{\mathbf{x}}_0}\mathcal{L}_{\mathrm{data}}$} \Comment{\textcolor{blue}{Data-consistency guidance}} 
\State $\kappa_t\leftarrow \min(k,t)$, \quad $\Delta\beta^{(\kappa_t)}\leftarrow \beta_t-\beta_{t-\kappa_t}$ \Comment{Avoid $t-k<0$} 
\State \textcolor{blue}{$\tilde{\mathbf{x}}\leftarrow \mathbf{x}_t-\Delta\beta^{(\kappa_t)}\mathbf{f}_t-\eta\mathbf{g}_t$} \Comment{\textcolor{blue}{Guided predictor step}} 
\State $\Gamma_t^{(\kappa_t)}\leftarrow \cos(\beta_{t-1})/\cos(\beta_{t-\kappa_t})$ \Comment{Variance-preserving scale} 
\State $\mathbf{x}_{t-1}\leftarrow \Gamma_t^{(\kappa_t)}\tilde{\mathbf{x}} +\sqrt{1-(\Gamma_t^{(\kappa_t)})^2}\boldsymbol{\epsilon}', \quad \boldsymbol{\epsilon}'\sim\mathcal{N}(0,\mathbf{I})$ \Comment{Corrector step} 
\State $\mathcal{X}\leftarrow \mathcal{X}\cup\{\mathbf{x}_{t-1}\}$ 
\EndFor 
\State \textbf{Return} $\mathcal{X}$ 
\end{algorithmic}
\end{algorithm}

TF-PRDiT solves inverse problems of the form
$\mathbf{y}=\mathcal{A}(\mathbf{x})+\boldsymbol{\xi}$,
where $\mathbf{x}$ is the unknown 3D volume, $\mathbf{y}$ is the measurement, $\mathcal{A}$ is a known forward operator, and $\boldsymbol{\xi}$ is measurement noise. The same formulation covers X-ray projection, downsampling, masking, and deblurring by changing $\mathcal{A}$. 

\vspace{3pt}
\noindent\textbf{Relation to DPS.}
Our sampler follows the same posterior-sampling perspective as DPS~\cite{chung2022diffusion}: an unconditional diffusion prior supplies the generative score, while a task-specific likelihood term enforces agreement with observed measurements. We therefore do not introduce a new posterior-guidance principle. Instead, TF-PRDiT specializes this principle for native 3D medical volumes by combining a voxel-level CT diffusion prior with differentiable volumetric forward operators. This specialization is important for sparse X-ray-to-CT because the forward model maps from a 3D volume to one or more 2D projections, so each additional X-ray can be incorporated by adding another projection-space residual to the same guidance loss.

\vspace{3pt}
\noindent\textbf{Frozen 3D diffusion prior.}
We use a pretrained PRDiT model~\cite{zhang2026pixellevel} as a frozen unconditional prior over chest CT volumes. PRDiT is a diffusion transformer trained directly in voxel space on LIDC-IDRI CT volumes~\cite{armato2011lung}, avoiding the compressed latent representation used by many latent diffusion models. This voxel-level formulation is important for medical reconstruction because fine anatomical boundaries and small structures may be weakened or lost during latent compression. Given a noisy volume $\mathbf{x}_t$ and timestep $t$, the frozen network jointly predicts the noise component and the corresponding clean-volume estimate: $(\hat{\boldsymbol{\epsilon}},\hat{\mathbf{x}}_0)=f_\theta(\mathbf{x}_t,t)$.
During downstream reconstruction, all parameters of $f_\theta$ remain fixed. Task adaptation is performed only through measurement-guided sampling, so the same prior can be reused across different view counts and inverse operators without retraining. 

\vspace{3pt}
\noindent\textbf{Predictor-corrector sampling.}
Using the cosine-sine parameterization~\cite{zhang2023improving}, let $\beta_t=\frac{\pi}{2}\frac{t}{T}$. The prior-induced update direction is 
\begin{equation}
\mathbf{f}_t=\sin(\beta_t)\hat{\mathbf{x}}_0-\cos(\beta_t)\hat{\boldsymbol{\epsilon}}.
\end{equation}
We use a $k$-step predictor,
\begin{equation}
\tilde{\mathbf{x}}=\mathbf{x}_t-\Delta\beta^{(k)}\mathbf{f}_t,\quad
\Delta\beta^{(k)}=\beta_t-\beta_{t-k},
\label{eq:k_scaled_predictor}
\end{equation}
which produces $\tilde{\mathbf{x}}\approx\mathbf{x}_{t-k}$ at noise level $t-k$. Because the predictor strides $k$ steps, the noise level of $\tilde{\mathbf{x}}$ is $\beta_{t-k}$, which differs from the target $\beta_{t-1}$ when $k>1$. A variance-preserving corrector rescales $\tilde{\mathbf{x}}$ and injects fresh noise to reach the correct noise level at timestep $t-1$:
\begin{equation}
\mathbf{x}_{t-1} = \Gamma_t^{(k)}\tilde{\mathbf{x}} + \sqrt{1-\bigl(\Gamma_t^{(k)}\bigr)^2}\,\boldsymbol{\epsilon}',\quad
\Gamma_t^{(k)} := \frac{\cos(\beta_{t-1})}{\cos(\beta_{t-k})},\quad \boldsymbol{\epsilon}'\sim\mathcal{N}(0,\mathbf{I}).
\label{eq:corrector}
\end{equation}
This keeps the marginal variance of $\mathbf{x}_{t-1}$ consistent with the forward process, preventing accumulated drift when $k>1$. Larger $k$ enables broader stochastic exploration, which is useful for ill-posed sparse-view reconstruction.

\vspace{3pt}
\noindent\textbf{Likelihood guidance on the denoised estimate.}
Directly matching measurements on the noisy state can produce unstable gradients. We instead guide sampling through the denoised estimate $\hat{\mathbf{x}}_0$, using the data-consistency loss
\begin{equation}
\mathcal{L}_{\mathrm{data}}=\|\mathcal{A}(\hat{\mathbf{x}}_0)-\mathbf{y}\|_2^2.
\label{eq:data_loss}
\end{equation}
The gradient $\nabla_{\hat{\mathbf{x}}_0}\mathcal{L}_{\mathrm{data}}$ is computed by backpropagating only through $\mathcal{A}$, treating $\hat{\mathbf{x}}_0$ as the free variable and holding the denoiser $f_\theta$ fixed. Following the DPS approximation~\cite{chung2022diffusion}, the Jacobian $\partial\hat{\mathbf{x}}_0/\partial\mathbf{x}_t$ is omitted for tractability; the resulting voxel-space gradient is applied directly as a correction to the $\mathbf{x}_t$ predictor update. For X-ray-to-CT, $\mathcal{A}$ is instantiated as a differentiable DRR projector using DiffDRR~\cite{gopalakrishnan2022fast}, allowing gradients to flow from projection space to voxels. With $M$ available X-rays, the loss becomes
\begin{equation}
\mathcal{L}_{\mathrm{xray}}=\sum_{i=1}^{M}\|\mathcal{P}_{g_i}(\hat{\mathbf{x}}_0)-\mathbf{y}_i\|_2^2,
\label{eq:multi_view_loss}
\end{equation}
where $\mathcal{P}_{g_i}$ denotes projection under view geometry $g_i$. Thus changing the number or geometry of views changes only the residual terms in the loss, not the network architecture or weights.

\vspace{3pt}
\noindent\textbf{Cosine-decay guidance.}
We use a time-dependent guidance scale $\eta_t=\eta_{\max}\cdot (1-\cos(\pi t/T))/2$.
At high-noise timestep, $\eta_t\approx\eta_{\max}$ provides strong measurement guidance to shape global structure. As $t\to1$, $\eta_t$ decays to zero, reducing over-correction and preserving fine anatomical details. This avoids the trade-off of fixed guidance, which can under-correct at high or over-correct at low noise.

\vspace{3pt}
Algorithm~\ref{alg} assembles the above components into the full conditional sampling procedure. At each timestep, the frozen prior predicts $(\hat{\boldsymbol{\epsilon}}, \hat{\mathbf{x}}_0)$, the predictor applies the prior-induced update together with measurement-consistency guidance, and the corrector restores the target noise level. Task adaptation is entirely controlled by $\mathcal{A}$ and $\mathbf{y}$, while $f_\theta$ remains fixed.

\section{Experiments}
\label{sec:exp}

\begin{table}[t]
\centering
\small
\setlength{\tabcolsep}{4pt}
\caption{Quantitative comparison of 1-view and 2-view X-ray-to-CT reconstruction on LIDC-IDRI. Results are mean\,$\pm$\,std over three random seeds, where available. \textsuperscript{\dag}indicates results reported in original papers. MSE is scaled by $10^3$.} \vspace{2pt}
\label{tab:final_results}
\begin{tabular}{lc|cccc}
\toprule
\textbf{Method} & \textbf{\#V} & \textbf{MSE}\,$\downarrow$ & \textbf{PSNR}\,$\uparrow$ & \textbf{SSIM}\,$\uparrow$ & \textbf{SNR}\,$\uparrow$ \\ \midrule
X2CT-GAN~\cite{ying2019x2ct}\textsuperscript{*} & 1 & \textbf{17.41} {\scriptsize $\pm$ 0.11} & \textbf{22.13} {\scriptsize $\pm$ 0.03} & \textbf{0.518} {\scriptsize $\pm$ 0.002} & \textbf{6.78} {\scriptsize $\pm$ 0.03} \\
X2CT-GAN~\cite{ying2019x2ct} & 1 & 18.91 {\scriptsize $\pm$ 0.41} & 21.75 {\scriptsize $\pm$ 0.10} & 0.509 {\scriptsize $\pm$ 0.008} & 6.39 {\scriptsize $\pm$ 0.10} \\
\rowcolor{gray!15}TF-PRDiT\,(ours) & 1 & 22.51 {\scriptsize $\pm$ 1.46} & 21.34 {\scriptsize $\pm$ 0.30} & 0.509 {\scriptsize $\pm$ 0.012} & 6.01 {\scriptsize $\pm$ 0.30} \\ \midrule
X2CT-GAN~\cite{ying2019x2ct}\textsuperscript{*} & 2 & 6.55 {\scriptsize $\pm$ 0.13} & 26.23 {\scriptsize $\pm$ 0.09} & 0.656 {\scriptsize $\pm$ 0.005} & 10.88 {\scriptsize $\pm$ 0.09} \\
X2CT-GAN~\cite{ying2019x2ct} & 2 & 6.80 {\scriptsize $\pm$ 0.07} & 26.05 {\scriptsize $\pm$ 0.05} & 0.647 {\scriptsize $\pm$ 0.004} & 10.70 {\scriptsize $\pm$ 0.05} \\
PerX2CT~\cite{kyung2023perspective}\textsuperscript{\dag} & 2 & \;--- & 27.45 {\scriptsize $\pm$ \,-----\,} & 0.732 {\scriptsize $\pm$ \;-----\,\;} & \!\quad--- \\
DiffuX2CT~\cite{liu2024diffux2ct}\textsuperscript{\dag} & 2 & \;--- & 26.35 {\scriptsize $\pm$ \,-----\,} & 0.687 {\scriptsize $\pm$ \;-----\,\;} & \!\quad--- \\
DX2CT~\cite{jeong2025dx2ct}\textsuperscript{\dag} & 2 & \;--- & 28.36 {\scriptsize $\pm$ \,-----\,} & 0.763 {\scriptsize $\pm$ \;-----\,\;} & \!\quad--- \\
\rowcolor{gray!15}TF-PRDiT\,(ours) & 2 & \textbf{3.65} {\scriptsize $\pm$ 0.12} & \textbf{29.06} {\scriptsize $\pm$ 0.12} & \textbf{0.767} {\scriptsize $\pm$ 0.002} & \textbf{13.72} {\scriptsize $\pm$ 0.12} \\ \bottomrule
\end{tabular}
{\scriptsize\par\textsuperscript{*}X-rays generated with Plastimatch~\cite{sharp2010plastimatch}; \textsuperscript{\dag}MSE/SNR \& std not reported; code not publ. available.}
\end{table}

\noindent\textbf{Setup.}
We evaluate on LIDC-IDRI~\cite{armato2011lung} using the X2CT-GAN~\cite{ying2019x2ct} split (916 train and 102 held-out test CTs), following its preprocessing protocol (CT intensities clamped to [0,2500] HU) for fair comparison. The frozen PRDiT prior is trained once exclusively on the 916 training cases to match this benchmark setting; the 102 test cases are held out entirely. TF-PRDiT itself remains training-free: no weights are updated for reconstruction, and all task adaptation happens at inference via measurement-guided sampling. Projections are generated with DiffDRR~\cite{gopalakrishnan2022fast}, and we report full-volume MSE, PSNR, SSIM~\cite{wang2004image}, and SNR.

\subsection{X-ray-to-CT Reconstruction}
\noindent\textbf{Baseline comparisons.}
Table~\ref{tab:final_results} summarizes quantitative results for both the single-view and biplanar X-ray-to-CT reconstruction tasks on the LIDC-IDRI dataset. The single-view setting is inherently ill-posed because a single projection provides insufficient depth information, leaving many anatomically plausible 3D volumes that are consistent with the same X-ray measurement. Despite relying solely on a frozen unconditional diffusion prior and requiring no task-specific training, TF-PRDiT achieves competitive performance with the supervised X2CT-GAN baseline, demonstrating that a strong generative prior can effectively regularize this severely under-constrained inverse problem.

\begin{figure*}[t]
    \centering
    \includegraphics[width=\textwidth]{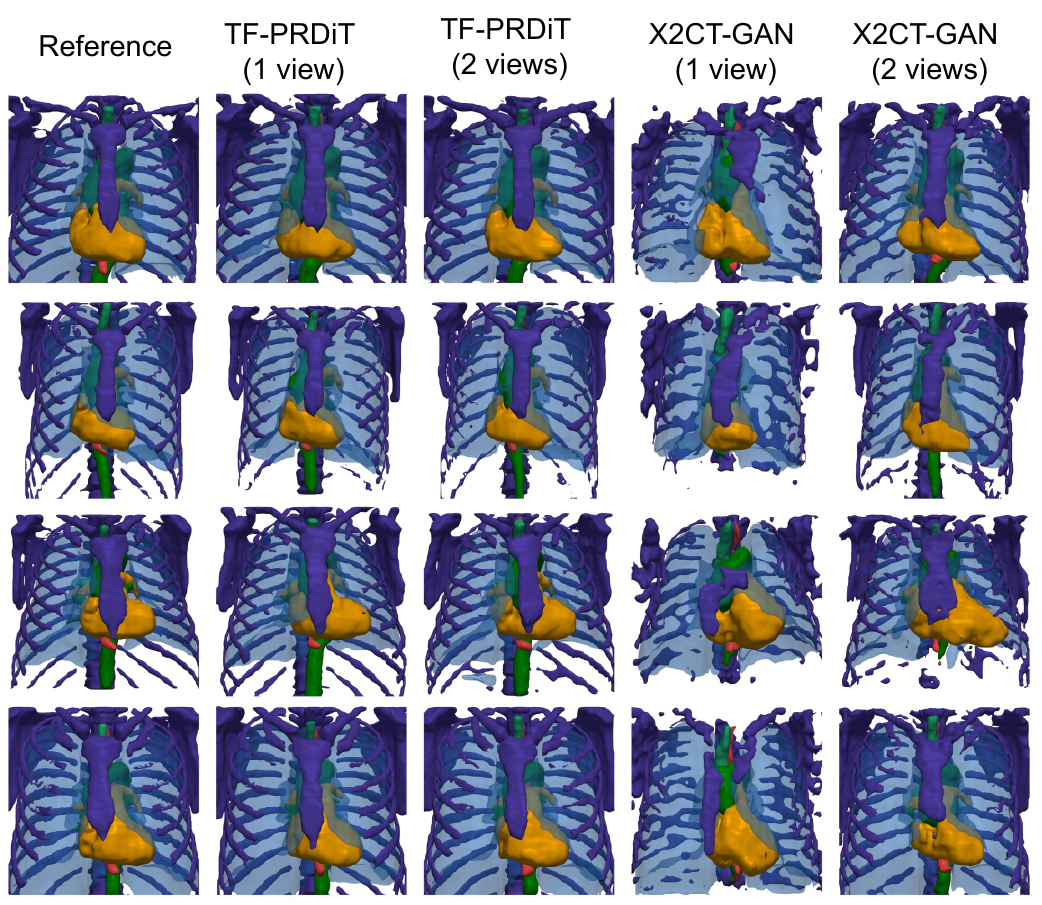}
    \caption{
    Organ-level qualitative comparison of reconstructed CT volumes on
    LIDC-IDRI. Anatomical regions are segmented using TotalSegmentator~\cite{wasserthal2023totalsegmentator}
    and rendered with consistent colors across the reference and reconstructed
    volumes. Compared with the single-view setting and X2CT-GAN, the biplanar
    TF-PRDiT reconstruction better preserves organ shape, spatial arrangement,
    and anatomical boundaries.
    }
    \label{fig:segment}
\end{figure*}

The advantage of TF-PRDiT becomes much more apparent in the biplanar setting. By incorporating two orthogonal X-ray projections, the geometric ambiguity is substantially reduced, allowing the diffusion prior to recover more accurate patient-specific anatomy. TF-PRDiT achieves the best performance across all quantitative metrics, outperforming X2CT-GAN, PerX2CT~\cite{kyung2023perspective}, DiffuX2CT~\cite{liu2024diffux2ct}, and the recent state-of-the-art DX2CT~\cite{jeong2025dx2ct}. In particular, TF-PRDiT improves the PSNR from 28.36~dB to 29.06~dB and increases SSIM from 0.763 to 0.767 compared with DX2CT, while simultaneously achieving the lowest MSE and highest SNR.

To further evaluate reconstruction fidelity beyond voxel-wise metrics, Table~\ref{tab:organ_mae} reports organ-wise Hounsfield Unit (HU) mean absolute error (MAE) computed within anatomical regions defined by TotalSegmentator~\cite{wasserthal2023totalsegmentator}. TF-PRDiT consistently reduces HU error across most organs and lowers the average organ-wise MAE from 179.3~HU to 118.7~HU in the biplanar setting. These results indicate that the reconstructed volumes preserve tissue-specific attenuation values more accurately, which is particularly important for quantitative clinical analysis.

Figure~\ref{fig:segment} complements the numerical evaluation with an organ-level qualitative comparison. The segmentation maps reveal that single-view reconstruction remains affected by depth ambiguity, which can lead to distorted organ geometry and inaccurate spatial relationships. With two orthogonal projections, TF-PRDiT recovers organ shapes and relative anatomical positions more faithfully, particularly around the
lungs, liver, kidneys, and central thoracic structures. Compared with X2CT-GAN, the reconstructed anatomy exhibits fewer fragmented regions and better agreement with the reference segmentation, supporting the improved organ-wise HU measurements reported in Table~\ref{tab:organ_mae}.

\begin{figure}[t]
  \centering
  \includegraphics[width=1.0\linewidth]{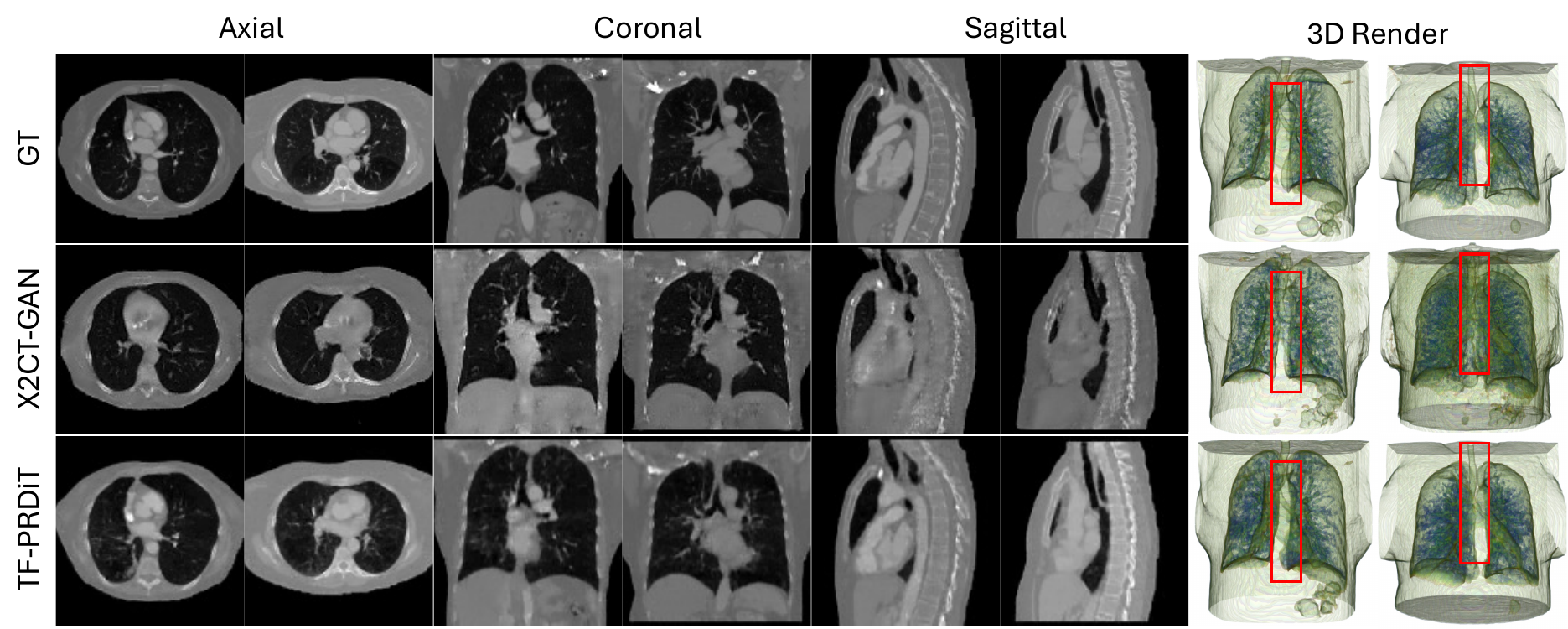}
  \caption{Comparison of bi-planar X-ray-to-CT reconstruction on LIDC-IDRI~\cite{armato2011lung}.}
  \label{fig:2views}
\end{figure}

Figure~\ref{fig:2views} provides qualitative comparisons. Compared with previous methods, TF-PRDiT reconstructs sharper anatomical boundaries, more faithful lung structures, and fewer streaking or over-smoothing artifacts across axial, coronal, and sagittal planes. The visual improvements closely align with the quantitative gains, demonstrating that the diffusion prior not only improves pixel-level reconstruction accuracy but also enhances anatomical realism.

\noindent\textbf{Scaling to arbitrary views.}
Table~\ref{tab:scalability_results} shows monotonic gains from 1 to 12 views all metrics (PSNR $21.34\to34.61$ dB, SSIM $0.509\to0.880$, average HU-MAE $279.3\to68.8$), with the sharpest jump occurring at 2 views where the second projection resolves depth ambiguity.
\begin{table}[t]
\centering
\footnotesize
\setlength{\tabcolsep}{5.2pt}
\renewcommand{\arraystretch}{1.02}
\caption{
Clinical-region-level HU fidelity evaluation.
HU-MAE is computed within TotalSegmentator~\cite{wasserthal2023totalsegmentator} defined organ masks.
Reported are mean $\pm$ std.
}\vspace{2pt}
\label{tab:organ_mae}
\begin{tabular}{lcccc}
\toprule
\multirow{2}{*}{\textbf{Structure}}
& \multicolumn{2}{c}{\textbf{Single-view MAE (HU) $\downarrow$}}
& \multicolumn{2}{c}{\textbf{Biplanar MAE (HU) $\downarrow$}} \\
\cmidrule(lr){2-3}
\cmidrule(lr){4-5}
& \textbf{Ours} & \textbf{X2CT}
& \textbf{Ours} & \textbf{X2CT} \\
\midrule
Lung      
& \ms{212.1}{81.5}  & \bms{172.4}{44.5}  
& \bms{101.6}{22.2} & \ms{137.7}{23.6} \\

Airway    
& \ms{522.4}{217.3} & \bms{423.4}{148.7} 
& \bms{179.3}{56.4} & \ms{283.7}{77.2} \\

Heart     
& \ms{171.0}{145.1} & \bms{131.3}{71.0}  
& \bms{49.5}{23.6}  & \ms{81.0}{37.3} \\

Vessels   
& \bms{186.3}{94.9} & \ms{203.9}{75.1}   
& \bms{69.7}{29.7}  & \ms{125.7}{40.7} \\

Bone      
& \ms{381.4}{129.8} & \bms{359.8}{73.5}  
& \bms{187.1}{37.4} & \ms{262.8}{40.7} \\

Esophagus 
& \bms{202.6}{86.8} & \ms{262.3}{88.6}   
& \bms{124.8}{47.5} & \ms{185.0}{60.2} \\
\midrule
\textbf{Mean}
& 279.3 & \textbf{258.9}
& \textbf{118.7} & 179.3 \\
\bottomrule
\end{tabular}
\label{tab:organ_mae}
\end{table}

\begin{table}[t]
\centering
\caption{Reconstruction quality vs. number of X-ray views for TF-PRDiT (frozen prior); average HU-MAE across \{lung, airway, heart, vessels, bone, esophagus\}.}\vspace{2pt}
\label{tab:scalability_results}
\setlength{\tabcolsep}{4pt}
\begin{tabular}{c|ccccc}
\toprule
\textbf{\#Views} & \textbf{MSE}\,$\downarrow$ & \textbf{PSNR}\,$\uparrow$ & \textbf{SSIM}\,$\uparrow$ & \textbf{SNR}\,$\uparrow$ & \textbf{HU-MAE}\,$\downarrow$ \\ \midrule
1 & 22.51 {\scriptsize $\pm$ 1.46} & 21.34 {\scriptsize $\pm$ 0.30} & 0.509 {\scriptsize $\pm$ 0.012} & 6.01 {\scriptsize $\pm$ 0.30} & 279.3 {\scriptsize $\pm$ 125.9}\\
2 & 3.65 {\scriptsize $\pm$ 0.12} & 29.06 {\scriptsize $\pm$ 0.12} & 0.767 {\scriptsize $\pm$ 0.002} & 13.72 {\scriptsize $\pm$ 0.12} & 118.7 {\scriptsize $\pm$ 36.1}\\
4 & 1.96 {\scriptsize $\pm$ 0.02} & 31.71 {\scriptsize $\pm$ 0.06} & 0.824 {\scriptsize $\pm$ 0.000} & 16.37 {\scriptsize $\pm$ 0.06} & 97.7 {\scriptsize $\pm$ 27.4}\\
6 & 1.39 {\scriptsize $\pm$ 0.02} & 33.20 {\scriptsize $\pm$ 0.07} & 0.852 {\scriptsize $\pm$ 0.000} & 17.86 {\scriptsize $\pm$ 0.07} & 81.7 {\scriptsize $\pm$ 19.0}\\
8 & 1.19 {\scriptsize $\pm$ 0.01} & 33.89 {\scriptsize $\pm$ 0.05} & 0.865 {\scriptsize $\pm$ 0.000} & 18.55 {\scriptsize $\pm$ 0.05} & 75.7 {\scriptsize $\pm$ 16.8}\\
12 & \textbf{1.02} {\scriptsize $\pm$ 0.02} & \textbf{34.61} {\scriptsize $\pm$ 0.07} & \textbf{0.880} {\scriptsize $\pm$ 0.000} & \textbf{19.27} {\scriptsize $\pm$ 0.07} & \textbf{68.8} {\scriptsize $\pm$ 14.3}\\ \bottomrule
\end{tabular}
\end{table}


\begin{figure}[t]
  \centering
  \includegraphics[width=0.9\linewidth]{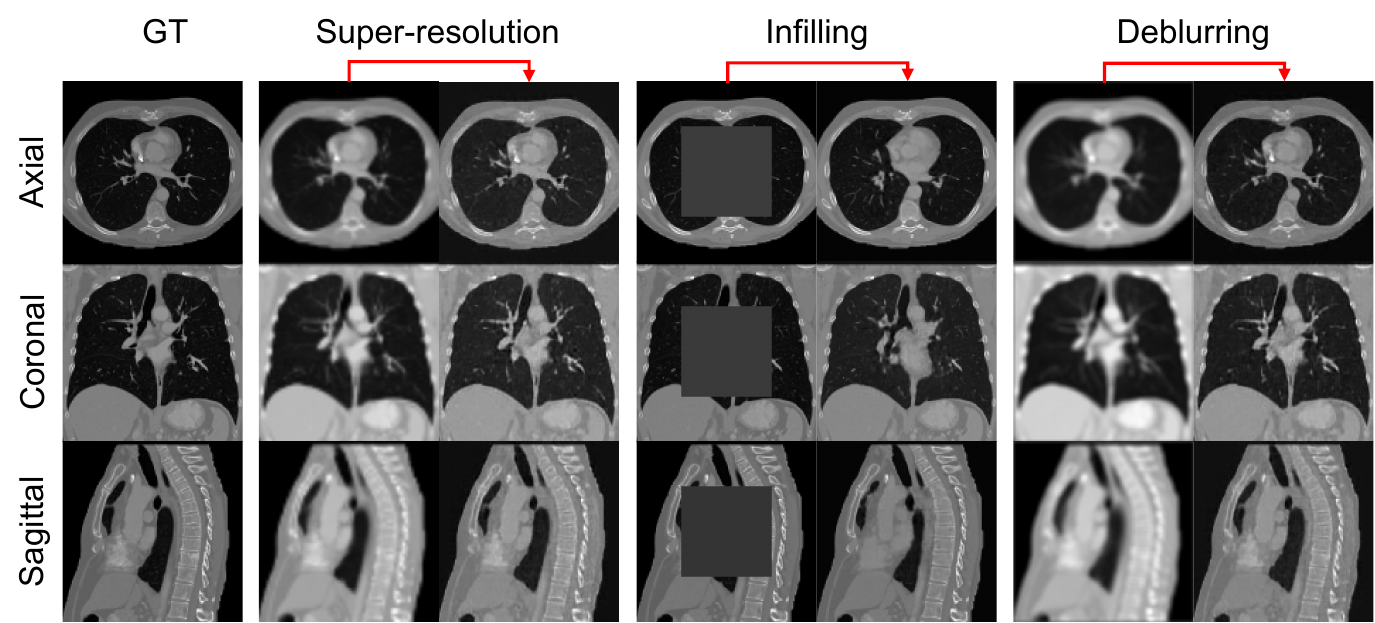}
  \caption{Qualitative results on additional 3D inverse problems. By replacing only the forward operator $\mathcal{A}$, TF-PRDiT handles super-resolution, volumetric infilling, and deblurring with the same frozen prior $f_\theta$.}
  \label{fig:inverse_problems}
\end{figure}

\subsection{Generalization and Ablation}

\noindent\textbf{Other inverse problems.}
Beyond X-ray-to-CT reconstruction, TF-PRDiT also generalizes to 3D super-resolution, volumetric infilling, and deblurring by only replacing the forward operator $\mathcal{A}$. 
As shown in Figure~\ref{fig:inverse_problems}, the same frozen diffusion prior restores missing or degraded structures across different degradation types without task-specific retraining. 
This supports the central motivation of TF-PRDiT: once a strong unconditional 3D prior is learned, different inverse problems can be addressed through measurement-consistent sampling rather than separate supervised models.

\begin{figure}[t]
  \centering
  \includegraphics[width=0.85\linewidth]{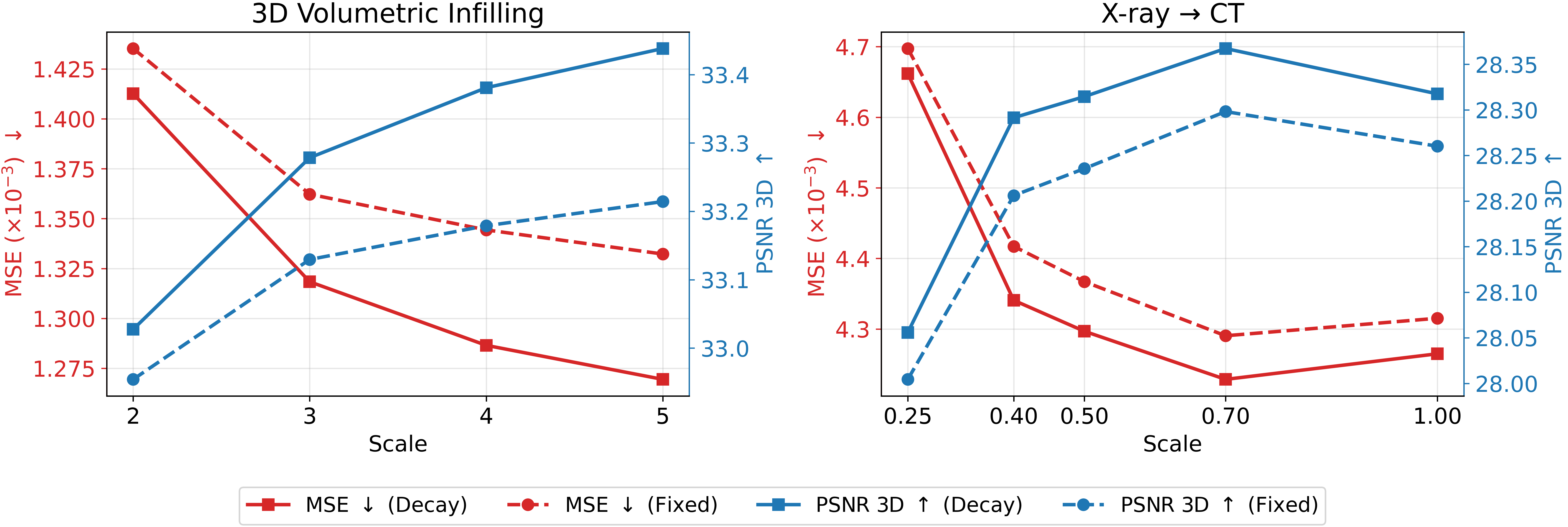}
  \caption{Ablation study on guidance schedule. Cosine decay vs.\ fixed guidance on 3D volumetric infilling (left) and 2-view X-ray-to-CT (right).}
  \label{fig:inp_x2ct}
\end{figure}

\vspace{3pt}
\noindent\textbf{Ablation.} 
We set $k=4$ for X-ray-to-CT and $k=2$ for $2\times$ super-resolution, reflecting the stronger ambiguity of sparse-view reconstruction. 
A larger $k$ promotes broader stochastic exploration, which helps resolve depth uncertainty when only limited projections are available, while a smaller $k$ is sufficient for the more constrained super-resolution task. 
Cosine-decayed guidance further outperforms fixed guidance on both tasks: strong early guidance corrects global structure, whereas decay near $t=1$ avoids over-correction and preserves fine anatomical details, as shown in Fig.~\ref{fig:inp_x2ct}.



\section{Conclusion}
\label{sec:con}

We presented TF-PRDiT, a training-free framework that converts a frozen voxel-level diffusion prior into a versatile solver for 3D medical inverse problems by swapping only the forward operator. Without retraining, it scales to 1--12 X-ray views and transfers to super-resolution, infilling, and deblurring.
Limitations include slower inference than feed-forward models and dependence on an accurate forward operator. 
Broader validation across anatomies, scanners, and clinical settings remains future work. Our code and pre-trained models are publicly available at \url{https://github.com/Fredy-Zhang/TF-PRDiT}.

\vspace{3pt}
\noindent\textbf{\ackname} This research was supported by The University of Melbourne’s Research Computing Services and the Petascale Campus Initiative. MH was supported by the Australian Research Council through grant DP230102775.

\vspace{3pt}
\noindent\textbf{\discintname}
The authors have no competing interests to declare that are relevant to the content of this article.

\FloatBarrier
\bibliographystyle{splncs04}
\bibliography{main}

\clearpage
\appendix

\section{Appendix}
\label{app:appendix}

The appendix is organized to separate implementation detail from empirical
evidence. This section records the common data, sampling, and evaluation
protocol. Section~\ref{app:operators} specifies how the same conditional
sampler is instantiated for each inverse problem. Section~\ref{app:architecture}
then documents the frozen 3D prior, after which
Secs.~\ref{app:predictor_ablation}--\ref{app:inverse} provide the predictor
ablation, arbitrary-view scaling results, and additional qualitative examples.

\subsection{Overview and Reproducibility Details}
\label{app:reproducibility}

\textbf{Data and preprocessing.} 
All experiments use LIDC-IDRI~\cite{armato2011lung} and follow the X2CT-GAN
split~\cite{ying2019x2ct}: 916 volumes are used to train the unconditional
prior and 102 held-out volumes are used for evaluation. CT intensities are
clamped to $[0,2500]$ HU and normalized before being passed to the $128^3$
voxel-space prior. Sparse X-ray measurements are generated with the
differentiable DiffDRR projector~\cite{gopalakrishnan2022fast}. No test volume,
projection, or paired reconstruction target is used to update the prior.

\vspace{3pt}
\noindent\textbf{Inference protocol.}
Every reconstruction starts from Gaussian noise and uses the frozen network
$f_\theta$ to predict both $\hat{\boldsymbol{\epsilon}}$ and
$\hat{\mathbf{x}}_0$. The data-consistency gradient is evaluated with respect
to $\hat{\mathbf{x}}_0$ and backpropagated only through the selected forward
operator $\mathcal{A}$; model parameters remain outside the optimization
graph. Guidance follows the cosine schedule
\[
\eta_t=\eta_{\max}\frac{1-\cos(\pi t/T)}{2},
\]
which emphasizes global measurement agreement early in sampling and reduces
over-correction as the sample approaches the clean volume. Unless otherwise
stated, the predictor multiplier is $k=4$ for sparse X-ray-to-CT and $k=2$ for
$\times2$ super-resolution, as selected by the ablation in
Table~\ref{tab:ablation_tasks}.

\vspace{3pt}
\noindent\textbf{Evaluation protocol.}
MSE, PSNR, SSIM~\cite{wang2004image}, and SNR are computed on the complete
reconstructed volume rather than on selected slices. Reported X-ray-to-CT
statistics are mean $\pm$ standard deviation over three random seeds. Visual
figures show matched axial, coronal, and sagittal locations, so the displayed
slices test cross-plane consistency rather than a single favorable plane.
Results copied from prior publications are identified as such in the main
paper because differences in projection generation, preprocessing, and metric
implementations prevent a strictly controlled comparison.

\FloatBarrier

\subsection{Forward-Operator Instantiations}
\label{app:operators}

The task-specific part of TF-PRDiT is fully described by the measurement
$\mathbf{y}$ and differentiable operator $\mathcal{A}$. In all cases the
sampler minimizes Eq.~\eqref{eq:data_loss}; only the residual used to construct
that loss changes. Table~\ref{tab:forward_operators} summarizes the four
instantiations evaluated in this work.

\begin{table}[htbp]
\centering
\small
\setlength{\tabcolsep}{4pt}
\renewcommand{\arraystretch}{1.12}
\caption{Task-specific forward operators. $\mathcal{P}_{g_i}$ is the DRR
projector at geometry $g_i$, $\mathcal{D}_s$ is downsampling by scale $s$,
$\mathbf{M}$ is a binary observation mask, and $\mathbf{K}$ is a blur kernel.}
\label{tab:forward_operators}
\vspace{2pt}
\begin{tabularx}{\linewidth}{@{}>{\raggedright\arraybackslash}p{0.23\linewidth}
  >{\raggedright\arraybackslash}p{0.25\linewidth}
  >{\raggedright\arraybackslash}X@{}}
\toprule
\textbf{Inverse problem} & \textbf{Forward model} & \textbf{Data-consistency residual} \\
\midrule
Sparse X-ray-to-CT & $\mathbf{y}_i=\mathcal{P}_{g_i}(\mathbf{x})$ &
$\sum_{i=1}^{M}\|\mathcal{P}_{g_i}(\hat{\mathbf{x}}_0)-\mathbf{y}_i\|_2^2$ \\
CT super-resolution & $\mathbf{y}=\mathcal{D}_s(\mathbf{x})$ &
$\|\mathcal{D}_s(\hat{\mathbf{x}}_0)-\mathbf{y}\|_2^2$ \\
Volumetric infilling & $\mathbf{y}=\mathbf{M}\odot\mathbf{x}$ &
$\|\mathbf{M}\odot\hat{\mathbf{x}}_0-\mathbf{y}\|_2^2$ \\
Medical deblurring & $\mathbf{y}=\mathbf{K}*\mathbf{x}$ &
$\|\mathbf{K}*\hat{\mathbf{x}}_0-\mathbf{y}\|_2^2$ \\
\bottomrule
\end{tabularx}
\end{table}

For X-ray-to-CT, adding a view appends one projection-space residual to the
sum; it does not change the network input channels, architecture, or weights.
For the remaining tasks, downsampling, masking, and convolution are linear and
differentiable, so their adjoints naturally map the measurement residual back
to the 3D voxel grid. This common interface is the practical reason a single
unconditional prior can be reused across all four problems. It also clarifies
the method's scope: reconstruction quality depends on the accuracy of
$\mathcal{A}$, and unmodelled acquisition geometry or corruption can produce a
biased guidance signal even when the generative prior is strong.

\FloatBarrier

\subsection{Enhanced PRDiT Image-Prediction Branch}
\label{app:architecture}

TF-PRDiT uses a frozen, unconditional PRDiT prior; the proposed training-free
contribution changes the sampling procedure and does not condition or fine-tune
the network. For completeness, we detail the implementation used in our
experiments and isolate its changes from the public PRDiT
architecture~\cite{zhang2026pixellevel}.

\vspace{3pt}
\noindent\textbf{Tokenization and shared timestep conditioning.}
For a noisy CT input defined as
$\mathbf{x}_t\in\mathbb{R}^{B\times1\times128\times128\times128}$, reflection-padded unfolding extracts overlapping $12^3$ voxel patches with stride 8 and padding 2. This produces a $16^3=4096$-token grid, with raw token
width $12^3=1728$. A 256-dimensional sinusoidal timestep encoding is mapped to
a shared 768-dimensional representation and then split into a
1728-dimensional coarse-conditioning vector and a 768-dimensional
refiner-conditioning vector.

\vspace{3pt}
\noindent\textbf{Local coarse denoiser.}
The first stage applies two timestep-modulated SwiGLU MLP blocks independently
to every raw patch token. A residual connection bypasses the two MLPs, after
which a linear head predicts an $8^3\times2=1024$-dimensional output patch.
This stage captures coarse local structure without global self-attention.

\vspace{3pt}
\noindent\textbf{Global residual refiner.}
The second stage embeds the same raw tokens using a nonlinear
$1728\!\rightarrow\!3072\!\rightarrow\!768$ path plus a linear
$1728\!\rightarrow\!768$ skip projection. Normalized 3D positional encodings
are added before Transformer blocks with multi-head self-attention, 64 channels
per head, a $4\times$ expansion MLP, and timestep-conditioned adaptive
normalization and residual gates. The final head predicts a 1024-dimensional
residual patch per token. Coarse and residual predictions are added in patch
space and rearranged into a dense two-channel $128^3$ output, corresponding to
noise and clean-image predictions. The prior is trained progressively: the
local denoiser is learned first and then frozen while the global refiner learns
the residual.

\begin{table}[H]
\centering
\small
\setlength{\tabcolsep}{5pt}
\renewcommand{\arraystretch}{1.08}
\caption{Changes restricted to the clean-image prediction branch
$\hat{\mathbf{x}}_0$. The noise-prediction branch $\hat{\boldsymbol{\epsilon}}$
and shared PRDiT backbone are unchanged.}
\label{tab:prdit_architecture}
\vspace{2pt}
\begin{tabularx}{\linewidth}{@{}>{\raggedright\arraybackslash}p{0.30\linewidth}
  >{\raggedright\arraybackslash}X>{\raggedright\arraybackslash}X@{}}
\toprule
\textbf{Image-branch component} & \textbf{Public PRDiT} & \textbf{Enhanced prior (ours)} \\
\midrule
Pre-normalization & Affine-free LayerNorm, $\epsilon=10^{-6}$ & Affine-free RMSNorm, $\epsilon=10^{-5}$ \\
AdaLN modulation output & SiLU--linear & SiLU--linear--RMSNorm, $\epsilon=10^{-5}$ \\
\bottomrule
\end{tabularx}
\end{table}
The modifications are therefore confined to normalization on the clean-image
prediction side; the noise-prediction side and shared local-to-global backbone
are retained. TF-PRDiT then supplies the orthogonal inference-time component:
the differentiable forward operator guides sampling while every parameter of
the prior remains fixed. Because these normalization changes are not ablated
individually, we do not attribute the reconstruction gain to either one in
isolation.

\FloatBarrier

\subsection{Predictor-Step Ablation}
\label{app:predictor_ablation}

Table~\ref{tab:ablation_tasks} evaluates the predictor-step multiplier $k$ on
sparse X-ray-to-CT reconstruction and $\times2$ CT super-resolution. For
X-ray-to-CT, increasing $k$ improves reconstruction quality up to $k=4$,
suggesting that the severely ill-posed projection-to-volume setting benefits
from broader exploration during sampling. In contrast, $\times2$
super-resolution achieves its lowest MSE at $k=2$, indicating that the more
constrained restoration task benefits from a conservative update. We therefore
use $k=4$ for sparse X-ray-to-CT and $k=2$ for $\times2$ super-resolution.

\begin{table}[htbp]
\vspace{-20pt}
\centering
\setlength{\tabcolsep}{4pt}
\caption{Ablation on predictor-step multiplier $k$. MSE is scaled by $10^3$.}
\label{tab:ablation_tasks}
\vspace{2pt}
\begin{tabular}{c|ccc|ccc}
\toprule
& \multicolumn{3}{c|}{\textbf{X-ray-to-CT}} & \multicolumn{3}{c}{\textbf{SR} ($\times2$)} \\
\textbf{$k$} & \textbf{MSE}$\downarrow$ & \textbf{PSNR}$\uparrow$ & \textbf{SSIM}$\uparrow$ & \textbf{MSE}$\downarrow$ & \textbf{PSNR}$\uparrow$ & \textbf{SSIM}$\uparrow$ \\ \midrule
2 & 4.09 & 28.49 & 0.726 & \textbf{0.42} & \textbf{38.12} & 0.914 \\
3 & 3.83 & 28.85 & 0.756 & 0.43 & 38.08 & \textbf{0.917} \\
4 & \textbf{3.65} & \textbf{29.06} & 0.767 & 0.46 & 37.86 & 0.915 \\
5 & 3.73 & 28.97 & \textbf{0.770} & 0.48 & 37.60 & 0.912 \\ \bottomrule
\end{tabular}
\end{table}

\subsection{Scalability Across X-ray Views}
\label{app:views}

\begin{figure}[H]
  \centering
  \includegraphics[width=0.9\linewidth]{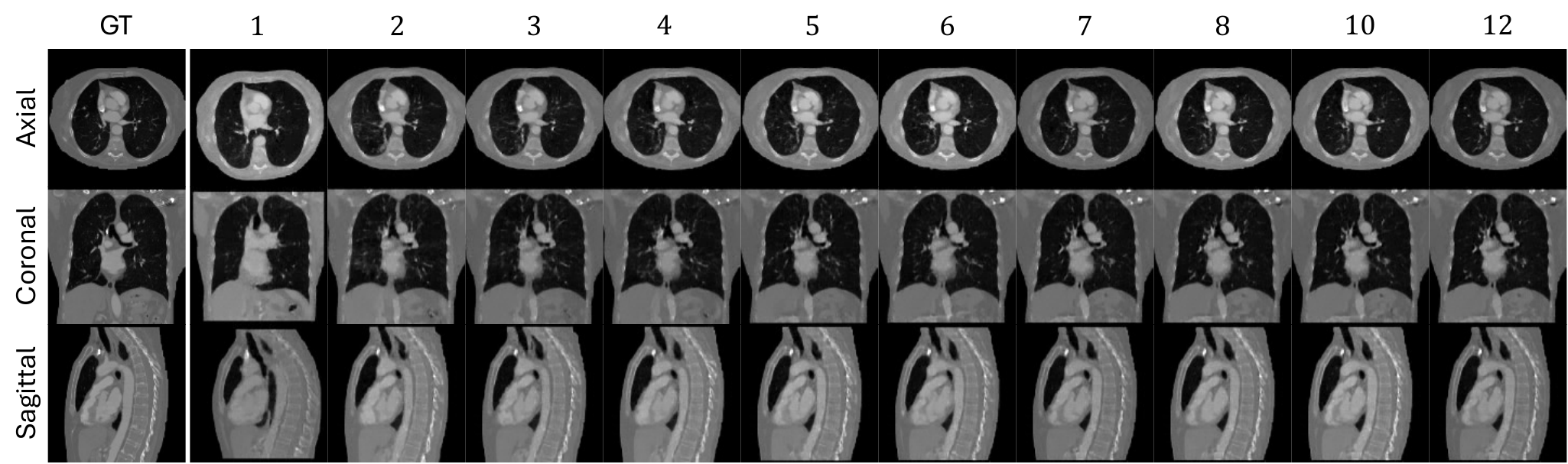}
  \caption{Qualitative scalability across 1--12 X-ray views. Axial, coronal,
  and sagittal slices are shown for each view count alongside ground truth (GT).}
  \label{fig:app_contviews}
\end{figure}

We evaluate whether the same frozen TF-PRDiT prior can accommodate different numbers of X-ray measurements without architectural modification or task-specific retraining. Table~\ref{tab:scalability_results_app} reports full-volume reconstruction metrics, while Fig.~\ref{fig:app_contviews} provides matched qualitative results as the number of input views increases from 1 to 12. A single projection is sufficient to recover the approximate global thoracic layout, but substantial depth ambiguity remains, leading to blurred boundaries and inaccurate fine structures. Introducing a second, approximately orthogonal view produces the largest improvement because it substantially reduces this ambiguity. Additional views continue to improve anatomical detail and cross-plane consistency, although the gains become progressively smaller as the reconstruction approaches saturation.

The qualitative progression in Fig.~\ref{fig:app_contviews} is consistent with the quantitative results in Table~\ref{tab:scalability_results_app}. PSNR increases from 21.34~dB with one view to 29.06~dB with two views, accompanied by a substantial SSIM improvement from 0.509 to 0.767. Performance continues to improve with additional measurements, reaching 34.61~dB PSNR and 0.880 SSIM with 12 views. Meanwhile, the average organ-level HU-MAE decreases from 279.30~HU to 68.77~HU, indicating that the additional projections improve not only global voxel-wise similarity but also the recovery of tissue-specific attenuation values.

\begin{table}[t]
\centering
\small
\setlength{\tabcolsep}{2.6pt}
\renewcommand{\arraystretch}{1.08}
\caption{Structure-specific HU-MAE as the number of input X-ray views increases.
Values are mean $\pm$ standard deviation across 102 test volumes. Lower values
indicate better preservation of clinically relevant attenuation values.}
\vspace{2pt}
\label{tab:organ_hu_mae_app}
\resizebox{\textwidth}{!}{
\begin{tabular}{c|cccccc}
\toprule
\textbf{\#Views} &
\textbf{Lung}$\downarrow$ &
\textbf{Airway}$\downarrow$ &
\textbf{Heart}$\downarrow$ &
\textbf{Great vessels}$\downarrow$ &
\textbf{Bone}$\downarrow$ &
\textbf{Esophagus}$\downarrow$ \\
\midrule
1  &
212.09 {\scriptsize $\pm$ 81.49} &
522.43 {\scriptsize $\pm$ 217.29} &
171.03 {\scriptsize $\pm$ 145.06} &
186.27 {\scriptsize $\pm$ 94.89} &
381.38 {\scriptsize $\pm$ 129.76} &
202.61 {\scriptsize $\pm$ 86.78} \\

2  &
101.57 {\scriptsize $\pm$ 22.19} &
179.30 {\scriptsize $\pm$ 56.38} &
49.48 {\scriptsize $\pm$ 23.63} &
69.71 {\scriptsize $\pm$ 29.72} &
187.11 {\scriptsize $\pm$ 37.36} &
124.76 {\scriptsize $\pm$ 47.48} \\

3  &
87.48 {\scriptsize $\pm$ 17.94} &
146.07 {\scriptsize $\pm$ 34.05} &
41.96 {\scriptsize $\pm$ 19.29} &
58.87 {\scriptsize $\pm$ 20.38} &
158.69 {\scriptsize $\pm$ 25.46} &
109.81 {\scriptsize $\pm$ 37.51} \\

4  &
82.42 {\scriptsize $\pm$ 17.05} &
145.47 {\scriptsize $\pm$ 43.81} &
41.65 {\scriptsize $\pm$ 19.83} &
58.20 {\scriptsize $\pm$ 21.13} &
148.53 {\scriptsize $\pm$ 23.09} &
110.12 {\scriptsize $\pm$ 39.45} \\

5  &
73.50 {\scriptsize $\pm$ 14.65} &
113.52 {\scriptsize $\pm$ 19.28} &
35.68 {\scriptsize $\pm$ 15.94} &
48.84 {\scriptsize $\pm$ 14.60} &
136.50 {\scriptsize $\pm$ 20.44} &
91.78 {\scriptsize $\pm$ 29.75} \\

6  &
71.57 {\scriptsize $\pm$ 14.17} &
112.41 {\scriptsize $\pm$ 20.30} &
34.91 {\scriptsize $\pm$ 15.77} &
48.80 {\scriptsize $\pm$ 15.23} &
130.00 {\scriptsize $\pm$ 19.39} &
92.29 {\scriptsize $\pm$ 29.10} \\

7  &
67.30 {\scriptsize $\pm$ 12.75} &
101.17 {\scriptsize $\pm$ 16.47} &
32.76 {\scriptsize $\pm$ 14.21} &
44.11 {\scriptsize $\pm$ 12.42} &
126.71 {\scriptsize $\pm$ 18.38} &
82.81 {\scriptsize $\pm$ 24.08} \\

8  &
66.44 {\scriptsize $\pm$ 12.54} &
103.06 {\scriptsize $\pm$ 17.67} &
32.49 {\scriptsize $\pm$ 14.14} &
45.26 {\scriptsize $\pm$ 13.35} &
122.38 {\scriptsize $\pm$ 18.08} &
84.37 {\scriptsize $\pm$ 25.09} \\

10 &
63.24 {\scriptsize $\pm$ 11.58} &
94.78 {\scriptsize $\pm$ 14.55} &
30.64 {\scriptsize $\pm$ 12.93} &
42.25 {\scriptsize $\pm$ 11.43} &
118.07 {\scriptsize $\pm$ 17.41} &
78.66 {\scriptsize $\pm$ 22.29} \\

12 &
\textbf{61.10} {\scriptsize $\pm$ 10.92} &
\textbf{91.04} {\scriptsize $\pm$ 14.76} &
\textbf{29.65} {\scriptsize $\pm$ 12.24} &
\textbf{40.60} {\scriptsize $\pm$ 10.86} &
\textbf{114.59} {\scriptsize $\pm$ 16.71} &
\textbf{75.66} {\scriptsize $\pm$ 20.41} \\
\bottomrule
\end{tabular}
}
\end{table}

\begin{table}[t]
\centering
\small
\setlength{\tabcolsep}{3.2pt}
\renewcommand{\arraystretch}{1.05}
\caption{Reconstruction quality as the number of input X-ray views increases. Values are mean $\pm$ standard deviation over three random seeds; MSE is scaled by $10^3$. Average HU-MAE is computed across five anatomical structures.}
\vspace{2pt}
\label{tab:scalability_results_app}
\begin{tabular}{c|ccccc}
\toprule
\textbf{\#Views} &
\textbf{MSE}$\downarrow$ &
\textbf{PSNR 3D}$\uparrow$ &
\textbf{SSIM 3D}$\uparrow$ &
\textbf{SNR}$\uparrow$ &
\textbf{HU-MAE}$\downarrow$ \\
\midrule
1  & 22.51 {\scriptsize $\pm$ 1.46} & 21.34 {\scriptsize $\pm$ 0.30} & 0.509 {\scriptsize $\pm$ 0.012} & 6.01 {\scriptsize $\pm$ 0.30} & 279.30 {\scriptsize $\pm$ 141.64} \\
2  & 3.65 {\scriptsize $\pm$ 0.12} & 29.06 {\scriptsize $\pm$ 0.12} & 0.767 {\scriptsize $\pm$ 0.002} & 13.72 {\scriptsize $\pm$ 0.12} & 118.65 {\scriptsize $\pm$ 56.35} \\
3  & 2.33 {\scriptsize $\pm$ 0.04} & 30.94 {\scriptsize $\pm$ 0.08} & 0.808 {\scriptsize $\pm$ 0.001} & 15.60 {\scriptsize $\pm$ 0.08} & 100.48 {\scriptsize $\pm$ 46.64} \\
4  & 1.96 {\scriptsize $\pm$ 0.02} & 31.71 {\scriptsize $\pm$ 0.06} & 0.824 {\scriptsize $\pm$ 0.000} & 16.37 {\scriptsize $\pm$ 0.06} & 97.73 {\scriptsize $\pm$ 44.62} \\
5  & 1.55 {\scriptsize $\pm$ 0.03} & 32.76 {\scriptsize $\pm$ 0.09} & 0.843 {\scriptsize $\pm$ 0.000} & 17.43 {\scriptsize $\pm$ 0.09} & 83.30 {\scriptsize $\pm$ 38.38} \\
6  & 1.39 {\scriptsize $\pm$ 0.02} & 33.20 {\scriptsize $\pm$ 0.07} & 0.852 {\scriptsize $\pm$ 0.000} & 17.86 {\scriptsize $\pm$ 0.07} & 81.66 {\scriptsize $\pm$ 36.78} \\
7  & 1.28 {\scriptsize $\pm$ 0.01} & 33.59 {\scriptsize $\pm$ 0.06} & 0.860 {\scriptsize $\pm$ 0.000} & 18.26 {\scriptsize $\pm$ 0.06} & 75.81 {\scriptsize $\pm$ 35.25} \\
8  & 1.19 {\scriptsize $\pm$ 0.01} & 33.89 {\scriptsize $\pm$ 0.05} & 0.865 {\scriptsize $\pm$ 0.000} & 18.55 {\scriptsize $\pm$ 0.05} & 75.67 {\scriptsize $\pm$ 34.30} \\
10 & 1.07 {\scriptsize $\pm$ 0.02} & 34.36 {\scriptsize $\pm$ 0.08} & 0.874 {\scriptsize $\pm$ 0.000} & 19.02 {\scriptsize $\pm$ 0.08} & 71.27 {\scriptsize $\pm$ 32.73} \\
12 & \textbf{1.02} {\scriptsize $\pm$ 0.02} &
\textbf{34.61} {\scriptsize $\pm$ 0.07} &
\textbf{0.880} {\scriptsize $\pm$ 0.000} &
\textbf{19.27} {\scriptsize $\pm$ 0.07} &
\textbf{68.77} {\scriptsize $\pm$ 31.70} \\
\bottomrule
\end{tabular}
\end{table}


\begin{figure}[t]
\centering
\includegraphics[width=\linewidth]{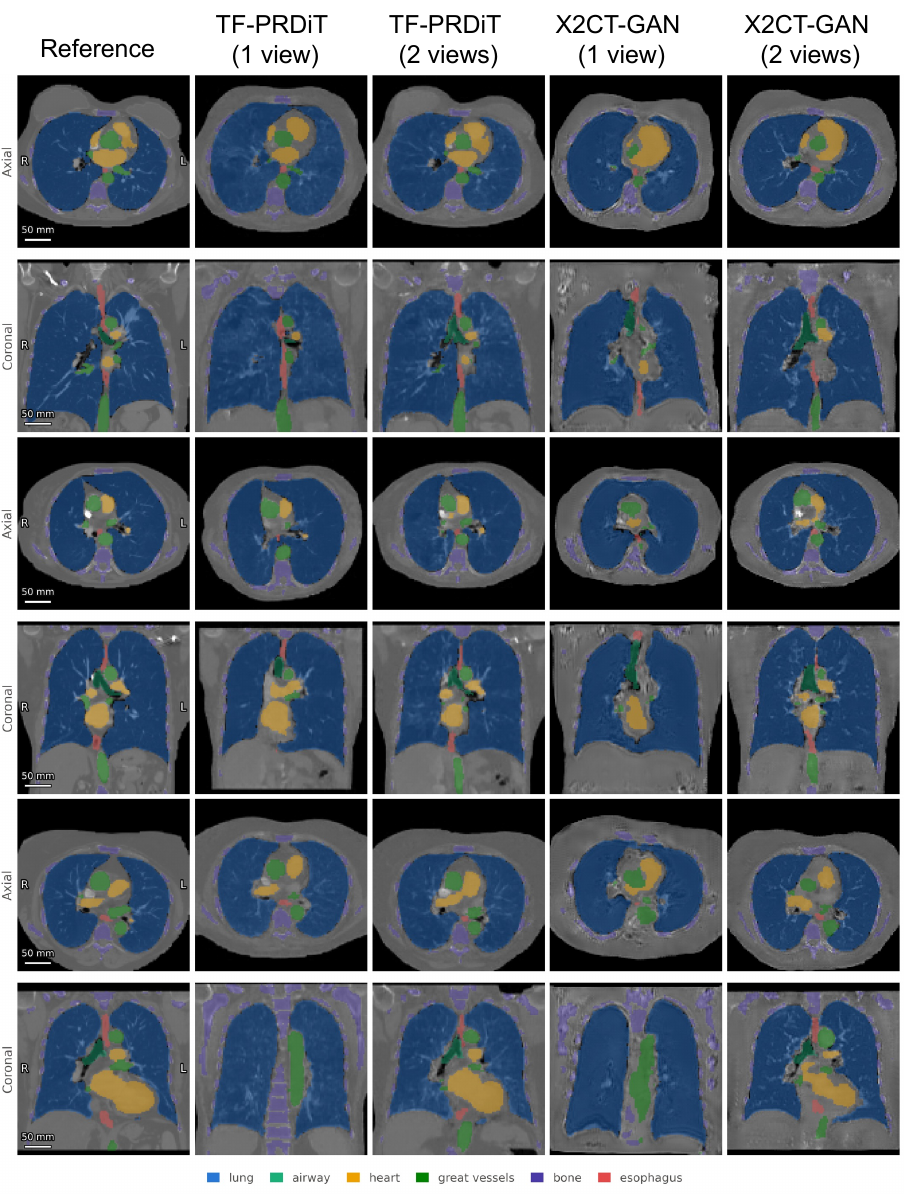}
\caption{Representative successful case. TF-PRDiT preserves the morphology and boundaries of the major anatomical structures more accurately than X2CT-GAN, particularly in the two-view setting.}
\label{fig:seg_success}
\end{figure}

\begin{figure}[t]
\centering
\includegraphics[width=\linewidth]{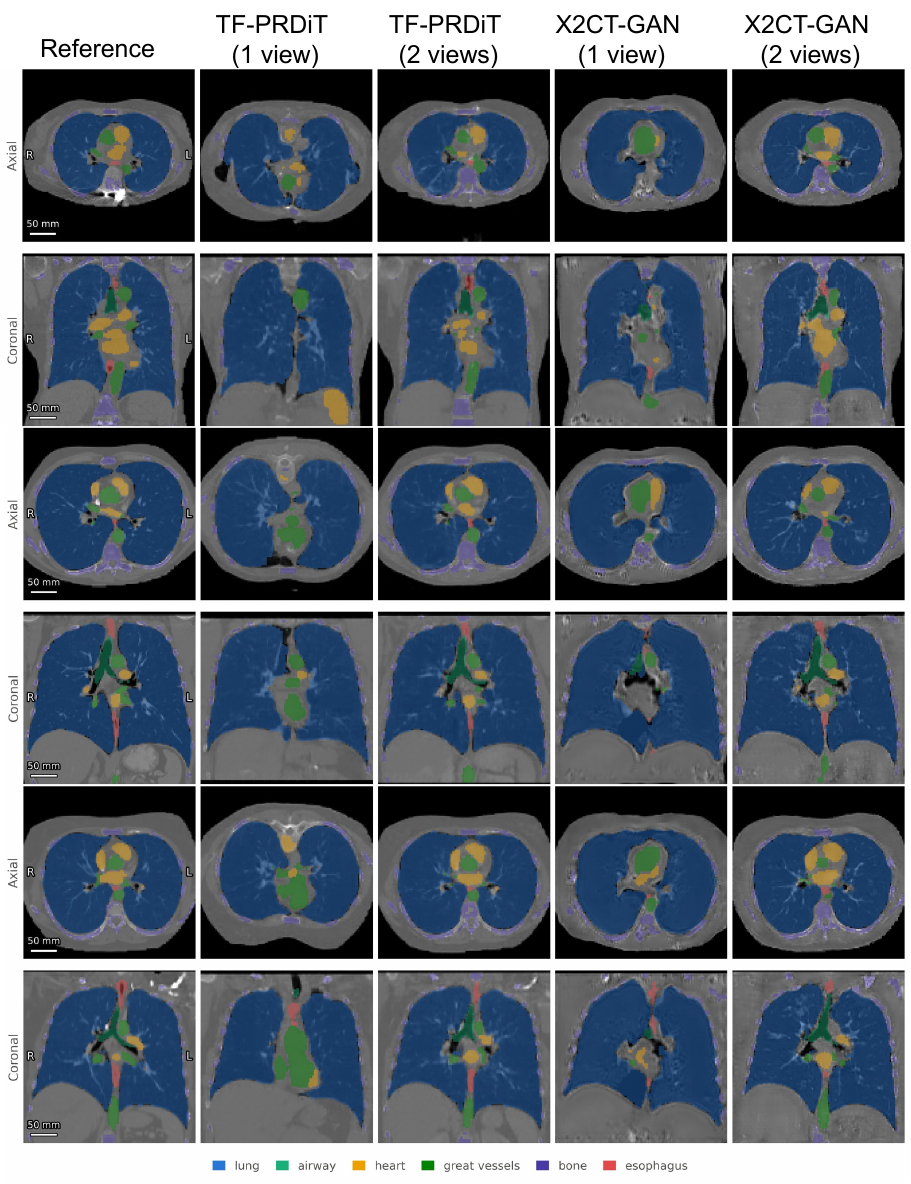}
\caption{Representative challenging 1-view case. Although TF-PRDiT recovers a plausible global thoracic layout, errors remain in small, thin, or low-contrast structures, illustrating the ambiguity of reconstruction from very sparse X-ray measurements.}
\label{fig:seg_challenging}
\end{figure}

\FloatBarrier
\subsection{Additional Qualitative Inverse-Problem Results}
\label{app:inverse}

\begin{figure}[H]
  \centering
  \begin{subfigure}[t]{\linewidth}
    \centering
    \includegraphics[width=\linewidth]{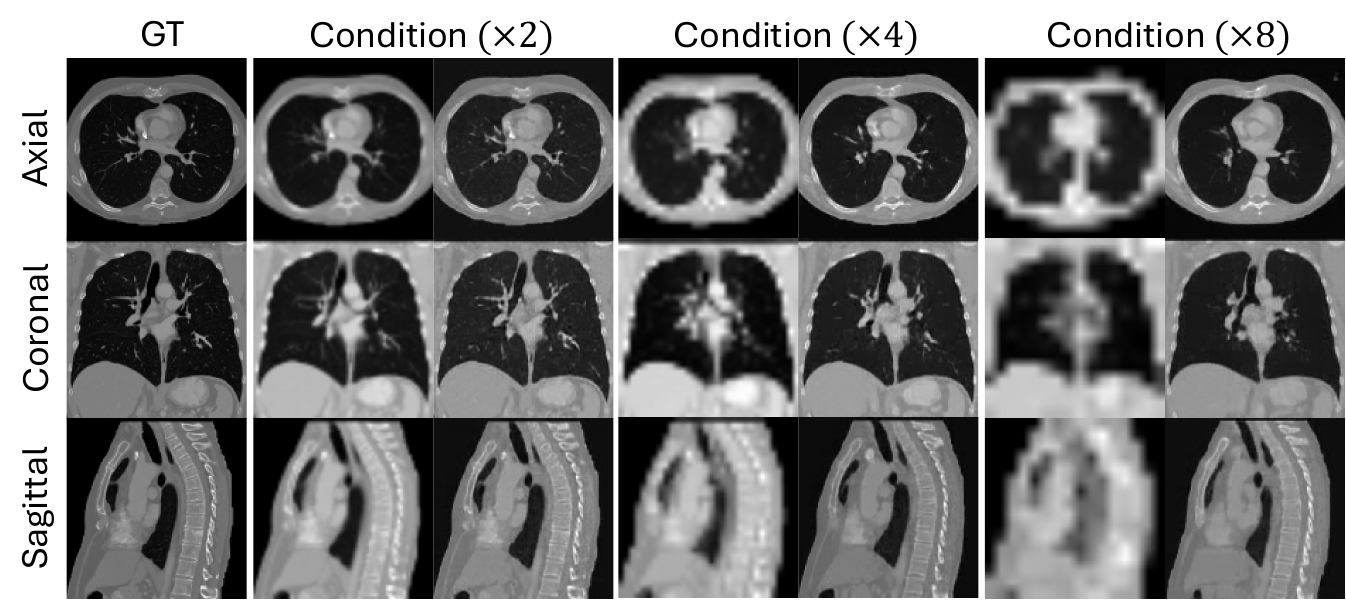}
    \caption{CT super-resolution at $\times2$, $\times4$, and $\times8$.}
  \end{subfigure}

  \vspace{0.8em}
  \begin{subfigure}[t]{0.48\linewidth}
    \centering
    \includegraphics[width=\linewidth]{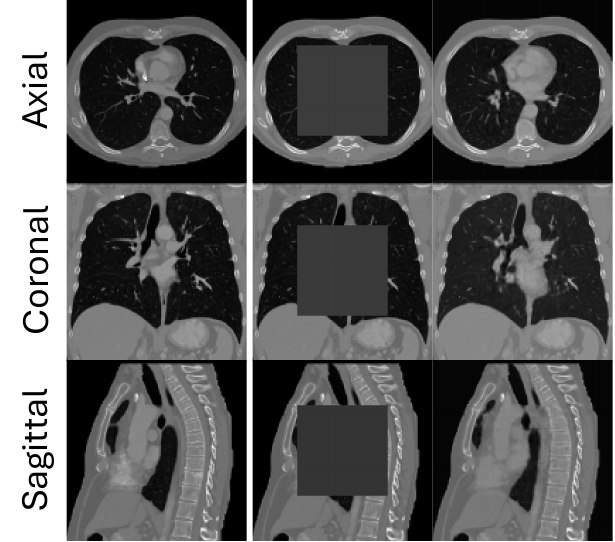}
    \caption{Volumetric infilling.}
  \end{subfigure}
  \hfill
  \begin{subfigure}[t]{0.48\linewidth}
    \centering
    \includegraphics[width=\linewidth]{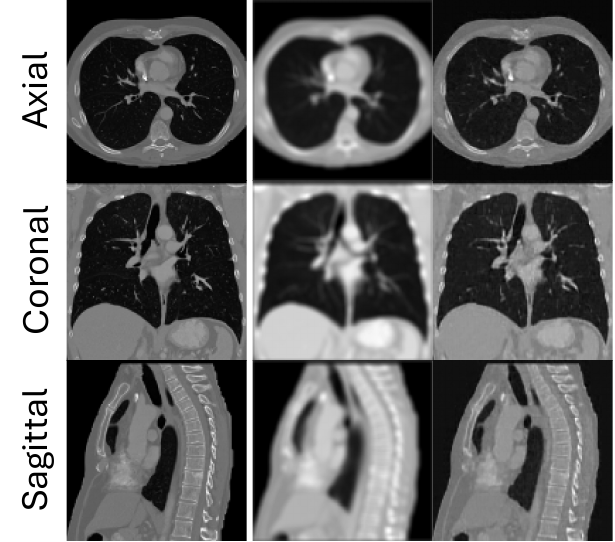}
    \caption{Medical image deblurring.}
  \end{subfigure}

  \caption{Additional inverse-problem results. Each panel shows ground truth,
  degraded input, and TF-PRDiT reconstruction.}
  \label{fig:app_inverse}
\end{figure}

Figure~\ref{fig:app_inverse} provides detailed visualizations for the additional 3D inverse problems discussed in Sec.~\ref{sec:exp}. In every case, only the forward operator $\mathcal{A}$ is changed; the diffusion prior $f_\theta$
remains frozen.

\FloatBarrier

\end{document}